\documentclass[12pt,preprint]{aastex}
\usepackage{hyperref}
\usepackage{cite}
\shorttitle{}

\shortauthors{Zhu et al.}

\def\degree{${}^{\circ}$}

\begin{document}

\title{Solar Energetic Particle Event Associated with the 2012 July 23 Extreme Solar Storm}

\author{Bei Zhu\altaffilmark{1,2}, Ying D. Liu\altaffilmark{1,2}, Janet G. Luhmann\altaffilmark{3},  Huidong Hu\altaffilmark{1,2}, Rui Wang\altaffilmark{1}, and Zhongwei Yang\altaffilmark{1}}

\altaffiltext{1}{State Key Laboratory of Space Weather, National Space 
Science Center, Chinese Academy of Sciences, Beijing 100190, China;
liuxying@spaceweather.ac.cn}

\altaffiltext{2}{University of Chinese Academy of Sciences, No.19A 
Yuquan Road, Beijing 100049, China}

\altaffiltext{3}{Space Sciences Laboratory, University of California, Berkeley, 
CA 94720, USA}

\begin{abstract}

We study the solar energetic particle (SEP) event associated with the 2012 July 23 extreme solar storm, for which STEREO and the spacecraft at L1 provide multi-point remote sensing and in situ observations. 
The extreme solar storm, with a superfast shock and extremely enhanced ejecta magnetic fields observed near 1 AU at STEREO A, was caused by the combination of successive coronal mass ejections (CMEs). 
Meanwhile, energetic particles were observed by STEREO and near-Earth spacecraft such as ACE and SOHO, suggestive of a wide longitudinal spread of the particles at 1 AU.
Combining the SEP observations with in situ plasma and magnetic field measurements we investigate the longitudinal distribution of the SEP event in connection with the associated shock and CMEs. 
Our results underscore the complex magnetic configuration of the inner heliosphere formed by solar eruptions.
The examinations of particle intensities, proton anisotropy distributions, element abundance ratios, magnetic connectivity and spectra also give important clues for the particle acceleration, transport and distribution.

\end{abstract}

\section{Introduction} 
With the launch of the \emph{Solar Terrestrial Relations Observatory} \citep[STEREO;][] {Kaiser08}, we now have multi-point measurements of SEPs including those from the L1 spacecraft such as \emph{Advanced Composition Explorer} \citep[ACE;][] {stone98} and \emph{SOlar and Heliospheric Observatory} \citep[SOHO;][]{Domingo95}.  
Both STEREO spacecraft follow a heliocentric orbit as the motion of the Earth in the ecliptic plane with their longitudinal separation increasing about 45\degree${ }$ per year. 
The configuration, which consists of the well separated STEREO and those near-Earth spacecraft including SOHO, ACE   and Wind, has proved to be a great platform to observe SEP events originating from any locations and associated processes from multiple vantage  points \citep[e.g.,][]{Liu11, Dresing12, Lario13, Richardson14, Cohen14, Gomez-Herrero15}. 

Several physical processes have been proposed to explain the wide longitudinal distribution of an SEP event in the interplanetary (IP) medium. 
A broad coronal or IP shock is one of the views \citep[e.g.,][]{Heras95, Lario98, Reames10b, Dresing12}. 
Particles observed at 1 AU in large SEP events are accelerated by CME-driven shocks and injected onto the magnetic field lines connecting the observers and the coronal shock \citep[e.g.,][]{Wild63, Cliver04, Zank07, Battarbee11, Cliver05, Kozarev15}, while the augular size of a wide coronal shock can extend up to 300\degree${ }$ \citep{Cliver95} and an IP shock at 1 AU can provide a large acceleration region with its longitudinal extent as large as 180\degree${ }$ \citep[e.g.,][]{Cane96, Liu08}.
On the other hand, 
EUV wave observed in the lower corona has been used as a proxy for the longitudinal extent of the CME during the initial expansion phase \citep[e.g.,][]{Torsti99, Rouillard12, Park13}. For the 2011 March 21 SEP event, \citet{Rouillard12} show an association between the longitudinal expansion of EUV wave and the longitudinal expansion of the SEP event at 1 AU. However, \citet{Prise14} suggest that the longitudinal spread of SEP event is related to CME expansion at a higher altitude in the corona than is represented by the expansion of EUV wave.
Alternatively, particle diffusion \citep[e.g.,][]{Reid64, Droge10, Dresing12, Costa13} and other processes like drifts \citep{Marsh13} and scattering are used to explain SEP transport in the corona and interplanetary medium.    
       
Another important factor that affects SEP transport and distribution is the complex magnetic configuration of the heliosphere and IP medium \citep[e.g.,][]{Richardson91, Liu11, Leske12, Masson12}.
For example, interplanetary CMEs (ICMEs) from previous solar eruptions can perturb the interplanetary magnetic field (IMF) structure and thus modify the particle travel path in the interplanetary medium. 
\citet{Park13} illustrate four SEP events that are probably influenced by preceding CMEs, which occur less than one day before the events. 
They suggest that the previous CME could influence the transport of the SEP event from the following CME. 
Actually, the preceding CME can occur earlier than one day. 
\citet{Liu14} conclude that the CME launched on 2012 July 19 from the same active region as the July 23 CMEs resulted in an IP medium with
low solar wind density and radial magnetic fields, which may have affected the transport of the SEP event from the July 23 solar event.
Information on the magnetic configuration of the heliosphere is needed to understand the particle transport and distribution.
Coronal magnetic field extrapolations based on photospheric magnetic field measurements can give a zeroth order characterization of the large-scale magnetic configuration \citep{Luhmann03}. 
In situ measurements of the solar wind plasma and magnetic field may provide a general context of the conditions through which SEPs propagate. 
Measurements of energetic particle anisotropy also provide important clues on the topology of the interplanetary magnetic fields  \citep[e.g.,][]{ Marsden87, Richardson91, Richardson94, Bieber02, Torsti04, Tan12, Leske12}.   
 
The solar storm on 2012 July 23 is of particular interest for space weather as it produced a superfast shock and extremely enhanced  magnetic fields at 1 AU \citep[e.g.,][]{Baker13, Ngwira13, Russell13, Liu14, Riley15, Temmer15}. 
\citet{Liu14} suggest that the extreme solar storm was caused by CME-CME interactions with the fast two CMEs separated by about 10-15 min. 
The merged CME structure had a speed of about 3050 km s${{}^{-1}}$ near the Sun and resulted in a solar wind speed of about 2246 km s${{}^{-1}}$ at 1 AU \citep{Liu14}. 
The magnetic field strength of the ejecta reached 109 nT at 1 AU. 
\citet{Temmer15} and \citet{Riley15} further examine the propagation of the shock and support the view of \citet{Liu14} that an earlier CME preconditioned the upstream solar wind for the propagation of the later eruptions.
In this paper we study the longitudinal distribution of the SEP event in connection with the shock and CMEs associated with the 2012 July 23 solar storm. 
As far as we are concerned, a detailed examination of the SEP event, in particular its connection with the shock and eruptions, has been lacking. 
\citet{Bain16} perform ENLIL MHD modeling to understand the shock connectivity associated with the SEP events during the whole 2012 July period.
Their simulations start from about 20 solar radii from the Sun and do not include flux rope magnetic fields in the ejecta.
Our work includes near-Sun magnetic mappings and considers the role of the ejecta in the SEP event, which will enhance the interpretation of this event.
First, through the examination of in situ plasma and magnetic field measurements, we give a scenario of the complex heliospheric configuration as a context to explain the SEP distribution. Second, we analyze the particle intensities, proton anisotropy distributions, element abundance ratios, magnetic connectivity and spectra and illustrate how these characteristics conform to the proposed scenario for the particle distribution. Finally, we summarize and discuss the results.

\section{Observations and Data Analysis}
\subsection{In Situ Plasma and Magnetic Field Measurements}
The extreme solar storm  originated from the active region 11520 (S15\degree W133\degree) and erupted at about 02:20 UT on 2012 July 23, while STEREO A and B were located at W121.3\degree${ }$ and E114.8\degree${ }$ with respect to the Earth respectively \citep{Liu14}. 
STEREO and near-Earth spacecraft constituted an all-around observing configuration for the July 23 event.
Figure 1 shows a summary of in situ plasma and magnetic field measurements from the STEREO and Wind spacecraft from 2012 July 23 to July 26. 
STEREO A observed a forward shock at 20:55 UT on July 23 followed by a complex ejecta composed of two ICMEs.
It is a typical eruption of twin CMEs discussed comprehensively by \citet{Liu14}. 
The launch times of the two CMEs were separated by about 10-15 min.    
The solar wind speed was up to 2246 km s${{}^{-1}}$ behind the shock, and the maximum magnetic field strength reached 109 nT inside the ejecta at 1 AU.
STEREO B observed an IP shock at 21:22 UT on July 23, which was followed by an ICME between 18:20 UT on July 24 and 12:00 UT on July 25. 
The ICME is identified mainly based on the enhanced magnetic field and rotation in the field components. 
No obvious shock and ICME signatures were observed at Wind during July 23-26. 
It is likely that the shock(s) and ICMEs missed the Earth. 

We reconstruct the ICME structure at STEREO B using a Grad-Shafranov (GS) technique \citep{hau99, hu02}, which has been validated by well-separated multi-spacecraft measurements \citep{Liu08, Mostl09}. 
The GS reconstruction gives a left-handed flux rope structure with an axis elevation angle of about ${-}$48\degree${ }$ and an azimuthal angle of about 93\degree${ }$ in RTN coordinates. 
As a contrast, the GS reconstruction gives a right-handed structure for each of the two ICMEs at STEREO A \citep{Liu14}. 
Clearly, the ICME observed at STEREO B is not the same event as observed at STEREO A.  
The arrival time of the shock at STEREO B is later than the shock at STEREO A about 27 minutes.
It is difficult to identify whether the shock at STEREO B is driven by the following ICME.
\citet{Bain16} suggest that the shock at STEREO B is likely associated with the small CME that left the Sun around 05:30 on July 19.
In the following discussions, the shock observed at STEREO B is denoted as ``the shock at STEREO B" and ``the shock" without specific statement refers to the shock observed at STEREO A.

Based on above measurements, we suggest a general context for the inner heliosphere, as illustrated in Figure 2. In this scenario, the ICMEs from different solar events were present simultaneously in the interplanetary medium during the period of July 23 event.
The ICMEs and shock(s) modified the global magnetic field configuration of the inner heliosphere, which may have played an important role in the transport and distribution of the SEPs produced by 2012 July 23 solar event. 
We will discuss the specific effects of the shock(s) and ICMEs on the SEP longitudinal distribution in the following sections. 
        
\subsection{Solar Energetic Particles}
Figure 3 shows the temporal evolution of proton intensity profiles observed at STEREO and near-Earth spacecraft (ACE and SOHO) during 2012 July 23-26. 
Proton intensities from STEREO are measured by (from top to bottom) the High-Energy Telescope \citep[HET;][]{Rosenvinge08}, Low-Energy Telescope \citep[LET;][] {Mewaldt08} and the Solar Electron and Proton Telescope \citep[SEPT;][] {Muller08}. 
The near-Earth proton intensities are measured by the Energetic Relativistic Nuclei and Electron instrument \citep[ERNE;][] {Torsti95} on board SOHO and the Electron, Proton, and Alpha Monitor \citep[EPAM;][]{Gold98} on ACE. 
\citet{Russell13} have discussed the correlation of the high energetic particle flux and the magnetic field, using the observations at STEREO A on 2012 July 23. 
Here we consider the transport and longitudinal spread of the SEP event. 
The positions of STEREO B, A and the Earth spanned a longitudinal extent of 245\degree${ }$ when the July 23 event occurred (Figure 2).
As shown in Figure 3, the SEP intensities increased clearly at all three spacecraft despite their wide separations. 
Meanwhile, intensity enhancements at all spacecraft had a long duration (above 3 days).
The peak intensities detected by STEREO A were around five orders of magnitude in high energy (${>}$10 MeV) bands higher than the background level.

The particle intensity-time profiles at STEREO A were consistent with the classic picture described by \citet{Cane88} and \citet{Reames99} for a SEP event near the central meridian.
\citet{Liu09} and \citet{Gopalswamy09} suggest that a CME-driven shock can form in the low corona.
\citet{Reames13} assumes that the strongest acceleration occurs near the nose of a shock that moves outward from the Sun with time.
However, the simulations by \citet{Liou14} indicate that the strongest part of the shock was not at the nose in the July 23 event.   
The prompt initial rise of proton intensities at STEREO A likely arose when STEREO A was connected to the west flank of the shock low in the corona.
Later, the observations showed a plateau behind the particle onset, a peak near the shock and a decrease in the ICME regions.
The plateau observed by STEREO A is probably caused by the `streaming limit' effect.
Protons from a shock near the Sun may generate Alfv{\'e}n waves that can scatter particles coming behind. Increasing the source proton intensity would enhance the wave growth, and the added scattering causes the proton intensity to level off at 1 AU \citep{Reames13}. 
Such a plateau has been studied in detail by \citet{Lee05}.
The protons near the shock can be the accelerated particles trapped by the the magnetic field fluctuations around the shock \citep{Lee83}. 
The intensity decrease in the ICME regions is attributed to a barrier formed by the strong magnetic fields inside the ICMEs.  

As shown in Figure 3, the proton intensities at STEREO B had an initial slow increase since the shock at STEREO B had arrived at 1 AU and the peak intensities occurred about 3 days after the shock at STEREO B passed (not shown here).
A classical eastern profile of \citet{Cane88} and \citet{Reames99} would start earlier and peak at or just beyond the shock passage.
The onset delay and long duration of the proton intensities at STEREO B may indicate that STEREO B was connected to the shock from behind as the shock passed 1 AU (see Figure 2).
Note that the proton intensities fluctuated as the observer entered the ICME, which is not the same ICMEs as  observed at STEREO A. 
This may indicate either an additional contributions to the SEP intensities from different solar activities or an effect on the SEP propagation of the July 23 event by the ICME detected at STEREO B.

Figure 4 shows the anisotropy distributions of 4-6 MeV protons from STEREO B during July 23-26.
A bidirectional proton streaming was visible for about 8 hours inside the ICME.
The sunward beam was broader than the beam from the antisunward direction.
\citet{Leske14} suggest that the anisotropic proton beams at STEREO B may be the result of magnetic mirroring and the shock may be the source of the antisunward streaming particles.
After the ICME, STEREO B observed largely isotropic proton distribution.
We tentatively suggest that those energetic particles after the ICME at STEREO B came from the acceleration of the shock and then were reflected by the magnetic fields within the ICME at STEREO B or scattered by the turbulence associated with this ICME.
Therefore, it is possible that the particle intensities at STEREO B primarily originated from the solar eruption of the July 23 event.

Observations near the Earth show that the intensity-time profiles of ${>}$1 MeV protons rose promptly and the proton intensities at ${>}$20 MeV energies declined over a long duration. 
The prompt rise was consistent with the picture of \citet{Cane88} and \citet{Reames99} for a western event relative to the observer, but the long decay was probably unexpected. 
The long decay at high energies may be related to the long time period over which the shock remained strong enough to accelerate particles.
Note that the Earth was on the eastern side of the twin CMEs and it missed the shock(s) and ICMEs at 1 AU. 
The particles observed near the Earth may have been accelerated by the shock close to the Sun and diffused longitudinally through the inner heliosphere to the footpoint of IMF lines connected to the Earth.

Figure 5 shows the abundance ratios for 13 elements (normalized to oxygen) obtained by integrating the intensities in the 12-33 MeV nucleon$^{-1}$ energy channel over the SEP event. 
\citet{Reames95} obtains the average element abundances of coronal SEPs in the energy channel of 5-12 MeV nucleon$^{-1}$, which are also shown in Figure 5.    
For C, Ar, Ca, Fe, and Ni, the abundance ratios relative to O at both STEREO and ACE spacecraft were lower than the average abundance ratios of the coronal component reported by \citet{Reames95}. 
The abundance ratios of other elements were similar to the reference ratios.     
In particular, the lower Fe$/$O ratios at STEREO and ACE spacecraft, compared with the average abundance ratio of the coronal component, likely indicates the origin of the SEP event, i.e., acceleration by the CME-driven shock.

\subsection{Magnetic Connectivity and Magneitic Field Line Length}
Figure 6 displays a GONG synoptic map at 5:54 UT on July 23 with open magnetic field lines simulated by a Potential Field Source Surface \citep[PFSS;][]{Schrijver03} model and connected to the ecliptic plane \href{http://gong.nso.edu/data/magmap/pfss.html}{(http://gong.nso.edu/data/magmap/pfss.html)}.
The longitudinal projections of both STEREO spacecraft and the Earth at 2.5 R$_{\odot}$ are calculated with a simple application of Parker spiral fields using the average solar wind speed measured in situ before the event onset time.
It is considered that the release times of particles at different observers are associated with the times of the coronal shock expanding to the magnetic footpoints of the spacecraft \citep[e.g.,][]{Rouillard12, Prise14}. 
As shown in Figure 6, the magnetic footpoint of STEREO A was the closest to the parent active region, and STEREO A was connected to the active region via a negative magnetic field. 
Particles may have arrived at the magnetic footpoint connected to STEREO A first and the magnetic footpoint connected to the Earth next as the CME-driven shock expanded.    
As we discussed above, the initial rise of the particle intensities at STEREO A and L1 was attributed to the acceleration by the shock near the Sun.
STEREO B was probably connected to the shock from behind when the particle intensities began to increase.  
Therefore, we suggest that the earliest particles observed at STEREO B were accelerated by the shock as it traveled beyond 1 AU, not near the Sun.  

Figure 7 shows the result of a velocity dispersion analysis (VDA) based on the onset times of protons at different energies observed by the LET and HET instruments onboard STEREO A.    
The blue straight line is a linear fit to the onset times at different energies with a simple expression \(t_{i}=t_{0}+{l}/{\nu}_{i}\), where ${t_{i}}$ is the onset time of particles, 
${{\nu}_{i}}$ is the particle velocity we evaluate using the geometrical mean of the minimum and maximum energies corresponding to the energy channel, $t_{0}$ is the particle release time on the Sun, and ${l}$ is the path length of the particles traveling from the Sun to the observer.
We determine the particle onset times using a Poisson-CUSUM method applied by \citet{Huttunen-Heikinmaa05}.
The estimate of the release time and field line length using the VDA method assumes: (1) particles at different energies are released simultaneously from the Sun and travel along a common path without energy dependency; (2) the earliest particles detected by the spacecraft are unscattered during the interplanetary propagation, and the particles with higher energies would arrive at 1 AU earlier than those with the lower energies \citep[e.g.,][]{Huttunen-Heikinmaa05, Lario14}.    
Using the proton onset times detected at STEREO A, the linear fit gives a path length of 1.46 ${\pm}$ 0.08 AU and a release time of 02:18 UT ${\pm}$ 4 minutes. Note that ${\sim}$8 minutes should be added to compare the proton release time with the electromagnetic observations, i.e., 02:26 UT ${\pm}$ 4 minutes. 
At this time the remote-sensing observations in STEREO B/COR1 show clearly twin CME structures \citep[see Figure 2(b) in][]{Liu14}.
The calculated path length is longer than the normal Parker spiral length (${\sim}$1.2 AU). 

We also calculate the release time and the path length of protons observed at L1 using the onset times detected by the ERNE instrument on SOHO, and obtain ${l \sim}$1.85 AU and ${t_{0} \sim}$05:50 UT. 
In the ENLIL modeling carried out by \citet{Bain16}, the time when the Earth was first connected to the simulated shock is 05:06 UT, which is earlier than the release time we calculated.  
Particle propagation may consist of two steps: traveling from the source region to the footpoint connected to the spacecraft as the coronal shock expands, and then propagating along the magnetic field lines to the spacecraft \citep[e.g.,][]{Park13, Richardson14}. 
The time interval between the solar eruption \citep[$\sim$02:20 UT;][]{Liu14} and the particle release time calculated with the SOHO observations is about 3.5 hours. 
This long delay may be a result of the time required to create a shock, to accelerate particles to high energies, and the time for the shock to reach the field lines connected to the Earth and the time for the accelerated particles to diffuse longitudinally.   
As we suggest above that the initial particles observed by STEREO B were accelerated by the shock beyond 1 AU, thus the VDA method was not performed on the STEREO B data.

\subsection{Energy Spectra}  
Figure 8 presents the proton energy spectra between 0.01 MeV and 100 MeV measured by HET, LET, SEPT on both STEREO spacecraft, by EPAM on ACE, and by ERNE on SOHO.
14-100 MeV proton intensity profiles at different locations are also shown in Figure 8, which indicate the time intervals where the spectra are derived.
The proton intensity profile observed by SOHO is given, along with a replotting of the STEREO proton intensities for ease of direct comparison, in panel (c).
In panels (d)-(f), proton energy spectra with different colors are identified as the background phase (blue), the rise phase (red), the peak phase (green), the ICME phase (purple) and the decay phase (black), respectively.
The ICME phase at STEREO A was taken during the second ICME and later than the peak phase while earlier at STEREO B.  
The ICME phase was absent at L1.
Panel (g) shows the spectra at different spacecraft when STEREO A and B entered the reservoir region.
The time interval was marked with the red bar in panel (c).

The temporal and spatial variations of the spectra are thought to be associated with the CME characteristics, the shock geometry and the observer's location \citep[e.g.,][]{Reames97, Sandroos07, Verkhoglyadova09, Verkhoglyadova10}. 
The spectra at both STEREO and near-Earth spacecraft showed major differences for different phases. 
The spectra at STEREO spacecraft hardened with time until the peak phase.
The shock(s) and ICMEs observed at STEREO A and B may have an effect on the particle spectra.
The particle reservoir began from ${\sim}$ 13:00 UT on July 26 and had a long duration at STEREO A and B.
Inside the reservoir region, the spectrum at STEREO A and B showed the similar shape and the spectrum near the Earth was around two order of magnitude lower than at STEREO A and B at ${<}$10 MeV energies.
As \citet{Reames97} identified, the invariance region of spectra begins after the shock passage but usually before the spacecraft passed through the ICME for the spacecraft located at the central and western flank of the shock.
During the July 23 event, The invariance region occurred about 1 day later when STEREO A and B passed through the ICMEs.
It seems that the particle reservoir began late as the shock passed the observer in a large gradual SEP event.

The shock arrival at STEREO B was 21:22 UT on July 23, and the ${>}$1 MeV proton intensities began to increase almost simultaneously.
It seems to support the view that the magnetic field lines connected STEREO B to the shock from behind and particles were transported along the magnetic field lines to STEREO B after the shock passed 1 AU. 
The spectra at STEREO B had little variability at high energies from the ICME phase to the peak phase.
STEREO B was probably connected to the western flank of the shock from behind (As shown in Figure 2).
The shock acceleration became weak with time as it propagated outwards.
Note that the time intervals of the peak phase and the decay phase were in the reservoir region.
The spectrum in the decay phase softened at ${>}$10 MeV energies compared with the spectrum in the peak phase. 
\citet{Reames13} suggests that the slow spectral steeping with time inside the reservoir region can be caused by the continuing acceleration by a weakening shock, preferential leakage of high energy particles or slower cross-field transport of lower energy particles.

For STEREO A, the spectrum showed sharp difference with time.
The spectrum in the rise phase formed the apparent suppression of the mid-range energy intensities (i.e., 0.5-10 MeV).
That is probably a streaming limit effect as discussed by \citet{Mewaldt13}.
The spectrum in the peak phase (around the shock) was much harder over the entire energy range than it was either before or afterward.
It seems that STEREO A passed through the stronger acceleration region of the shock.
Compared with the spectrum in the peak phase, the spectrum in the ICME phase kept a similar shape but was reduced by about two orders of magnitude at all energies.
The strong magnetic fields inside the ICMEs would form a barrier for particles.
The spectrum in the decay phase, which entered the reservoir region, declined at all energies and to the background level at ${<}$4 MeV energies.

For the spectrum near the Earth, the spectrum was invariant at ${<}$5 MeV energies in the rise phase and declined at ${<}$0.4 MeV energies in the peak phase compared with the spectrum in the background phase.
Note that the time interval of the peak phase near the Earth was earlier than the time that the shock passed STEREO A.
The Earth may have poor connection with the shock before it arrived at 1 AU.
As the spectra presented in panel (g), the spectrum near the Earth did not have the uniform spectral shape such as the spectra at STEREO A and B.
It seems that the Earth missed the reservoir region.
The spectrum in the decay phase declined at all energies.

\section{Conclusions and Discussion}
We have investigated the large SEP event associated with the 2012 July 23 extreme storm, which produced a superfast shock and extremely enhanced ejecta magnetic fields at STEREO A. 
Our analyses of the particle intensities, proton anisotropy distributions, element abundance ratios, magnetic connectivity and spectra provide important information on the origin, transport and longitudinal distribution of the large SEP event.
 
STEREO A and B, which were separated by about 124\degree , detected the shock(s) and ICMEs during the period of the July 23 event.
STEREO A observed the shock at 20:55 UT on July 23 followed by a complex ejecta composed of twin CMEs \citep{Liu14}. 
STEREO B observed an IP shock at 21:22 UT on July 23 with a following ICME between 18:20 UT on July 24 and 12:00 UT on July 25.
The GS reconstruction gives a left-handed structure for the ICME at STEREO B, wich is different from those at STEREO A \citep{Liu14}.
Therefore, we suggest that the ICME at STEREO B is not the same as the ICMEs at STEREO A.
It is unclear whether the shock observed at STEREO B is the same shock as observed at STEREO A.
The shock(s) and ICMEs may play an important role in the transport and longitudinal distribution of the SEP event.

Enhanced particle intensities were observed and had a long duration at both STEREO and near-Earth spacecraft, indicative of a wide longitudinal spread of SEPs. 
The Fe/O ratios at all three spacecraft were lower than the referenced abundance ratio of coronal component reported by \citet{Reames95}.
Therefore, the wide distributed SEP event on July 23 may largely originate from the acceleration by the CME-driven shock. 
STEREO A observed a typical SEP event with a source near the central meridian. 
The particle intensities rose rapidly after the solar eruption, and the peak intensities of ${>}$ 10 MeV particles observed at STEREO A were around five orders of magnitude higher than the background level. 
After the prompt initial rise of the particle intensities, a plateau behind the particle onset, a peak near the shock and a decrease inside the ICMEs were observed sequentially.
These can be interpreted as follows:
the earliest SEPs observed at STEREO A were accelerated by the shock near the Sun (initial rise), and could generate Alfv{\'e}n waves that would scatter the particles coming behind (plateau or streaming limit);
the magnetic field fluctuations near the shock could trap the SEPs accelerated by the IP shock (peak);
the strong magnetic fields inside the ICMEs could form the barrier for the SEPs (decrease).
The spectra in different phases display different characteristics of particle acceleration and propagation.
For STEREO A, the spectrum in the rise phase presented the apparent intensity suppression between 0.5-10 MeV energy range, caused by a streaming limit effect.
The spectrum in the peak phase was much harder at all energies compared with the spectra in other phases.
These spectra indicate that STEREO A passed through the stronger acceleration region of the shock.
The spectrum in the ICME phase kept a similar shape but intensities at all energies had about two orders of magnitude decrease than the spectrum in the peak phase.
Note that the decay phase at STEREO A was in the reservoir region.
Compared with the spectrum in the background phase, the hardened spectrum in the decay phase at ${>}$ 4 MeV energies may be accounted for by the property of the reservoir region.

STEREO B observed slow particle intensity enhancements since the shock at STEREO B arrived at 1 AU. 
The majority particles at STEREO B may be accelerated by the shock beyond 1 AU and then reach STEREO B along the IMF lines which connected the shock from behind.
Inside the ICME at STEREO B, we observed fluctuations in the particle intensities and obvious anisotropic proton distributions from ${\sim}$18:20 UT on July 24 to ${\sim}$3:00 UT on July 25.  
\citet{Leske14} suggest that the anisotropic flows may arise from the magnetic mirroring and the shock may be the source of the antisunward flow.
The following proton distribution became isotropic and proton intensities increase until STEREO B entered the reservoir region on July 26.
We suggest that the particle intensities at STEREO B primarily originated from the July 23 solar event.
The spectra at STEREO B kept similar shape in the ICME phase and the peak phase at ${>}$ 5 MeV energies, indicative of a connection between STEREO B and the weak acceleration region of the shock.
Note that the time intervals of the peak phase and the decay phase were in the reservoir region.
The steepened spectrum at ${>}$ 10 MeV energies in the decay phase may be attributed to continuing acceleration by a weakening shock, preferential leakage of high energy particles or slower cross-field transport of lower energy particles \citep{Reames13}.

The particle intensities observed by SOHO had a prompt rise at ${>}$1 MeV energies and a long duration at ${>}$20 MeV energies.
The long duration at high energies may be associated with the long time period of the strong acceleration of the shock.
Compared with the spectrum near the Earth in the background phase, the spectrum in the rise phase had little variability at ${<}$5 MeV energies and the spectrum in the peak phase declined at ${<}$0.4 MeV energies.
The Earth missed the shock(s) and ICMEs at 1 AU. 
It indicates the poor connection between the shock and the Earth.
The spectrum near the Earth in the decay phase was lower at all energies, where STEREO A and B entered the reservoir region.

We calculate the longitudinal projections of STEREO A, B and the Earth at 2.5 R$_{\odot}$ using a Parker spiral field.
The result shows that the magnetic footpoint of STEREO A was the closest to the active region.
SEPs accelerated by the shock near the Sun can be injected onto the open magnetic field lines connecting to STEREO A.
We also determine the particle release time at STEREO A using a VDA method. 
The linear fit gives a release time of 02:18 UT ${\pm}$ 4 minutes (02:26 UT ${\pm}$ 4 minutes, adding 8 minutes to compare with the remote-sensing observations).
The CMEs were clearly observed in STEREO B/COR1 at 02:26 UT (see Figure 2(b) in \citet{Liu14}).
It likely suggests that the earliest particles arriving at STEREO A were accelerated by the shock formed in the lower corona.
The magnetic footpoint of the Earth was relatively far from the active region.
The release time of the particles observed by SOHO is ${\sim}$05:50 UT.
It is about 3.5 hours late for the particle release since the solar eruption (${\sim}$ 02:20 UT).
The delay of the particle release may be attributed to the time to create a shock, to accelerate particles to high energies and the time for the shock to reach the field lines connected to the Earth and the time for the accelerated particles to diffuse to the footpoint of the IMF lines connected to the Earth.

\acknowledgments The research was supported by the Recruitment Program of Global Experts of China, NSFC under grant 41374173 and the Specialized Research Fund for State Key Laboratories of China. We acknowledge the use of data from STEREO, GONG, Wind, ACE and SOHO.

\newpage
\bibliography{ms_Jul23}
\bibliographystyle{apj}

\begin{figure}[!htb]
\centering
\begin{tabular}{ccc} 
\includegraphics[width=12.5pc]{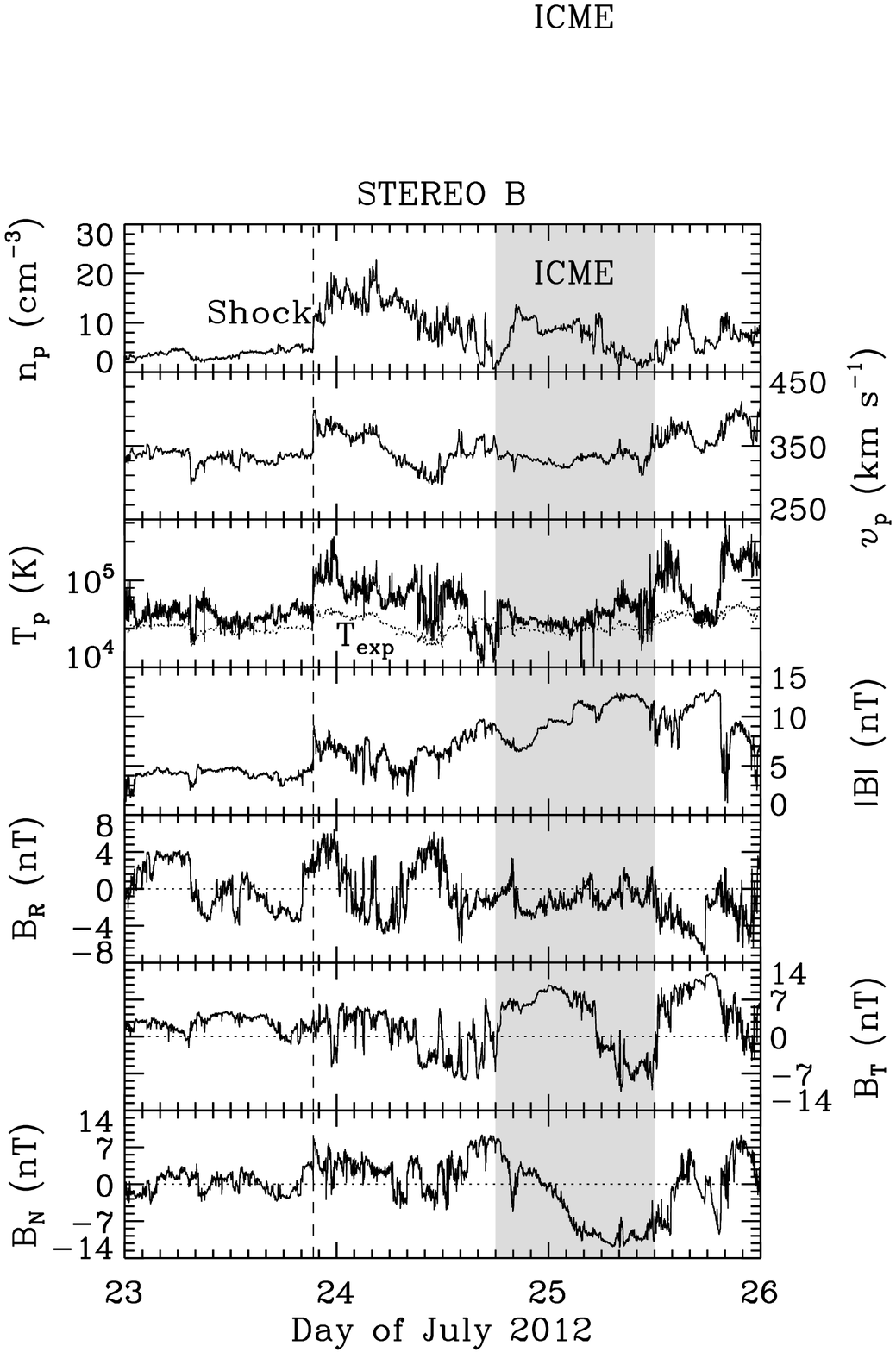}
\includegraphics[width=12.5pc]{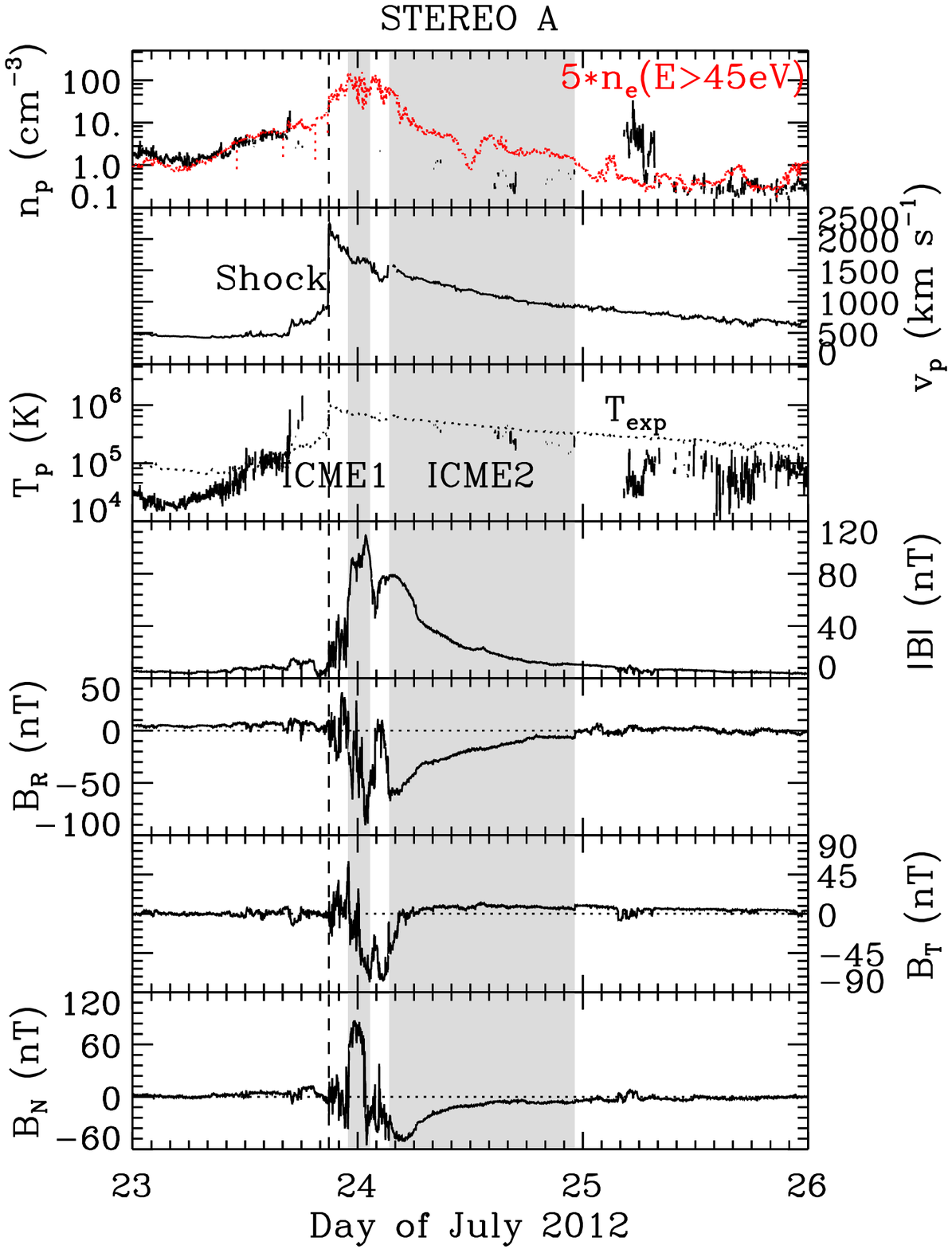}
\includegraphics[width=12.5pc]{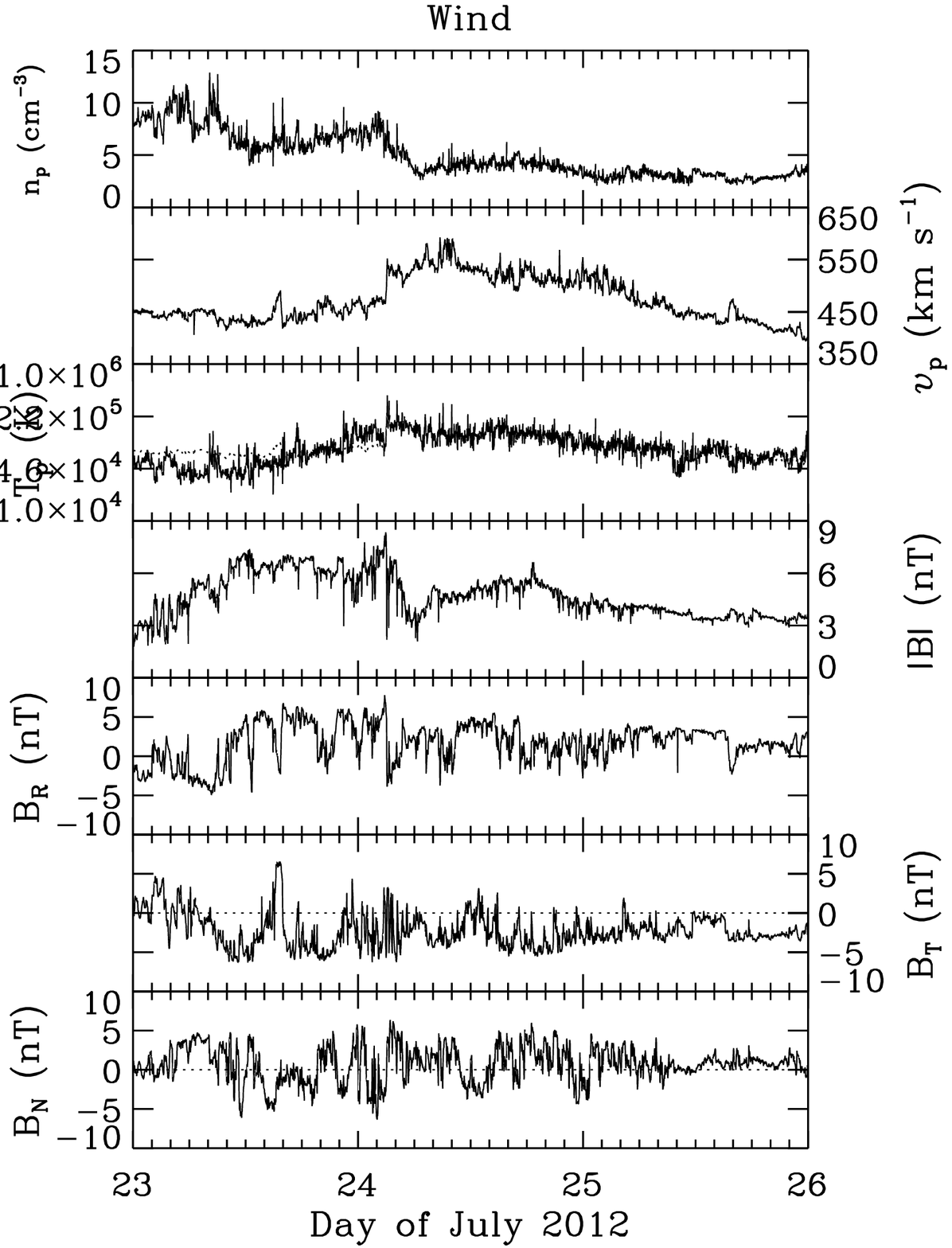}
\end{tabular}
\caption{In situ plasma and magnetic field measurements at STEREO B (left), STEREO A (middle; after \citet{Liu14}) and Wind (right). 
The panels from top to bottom show the proton density, bulk speed, proton temperature, magnetic field strength and components. 
ICME intervals are indicated by the shaded regions, and the observed shocks are marked by the dashed lines.
For STEREO A, the red curve in the top panel represents the number density (multiplied by a factor of 5) of electrons with energies above 45 eV.  
The dotted curve in third panel denotes the expected proton temperature calculated from the observed speed \citep{Lopez86}.}
\end{figure}

\begin{figure}[!htb]
\centering
\noindent\includegraphics[width=30pc]{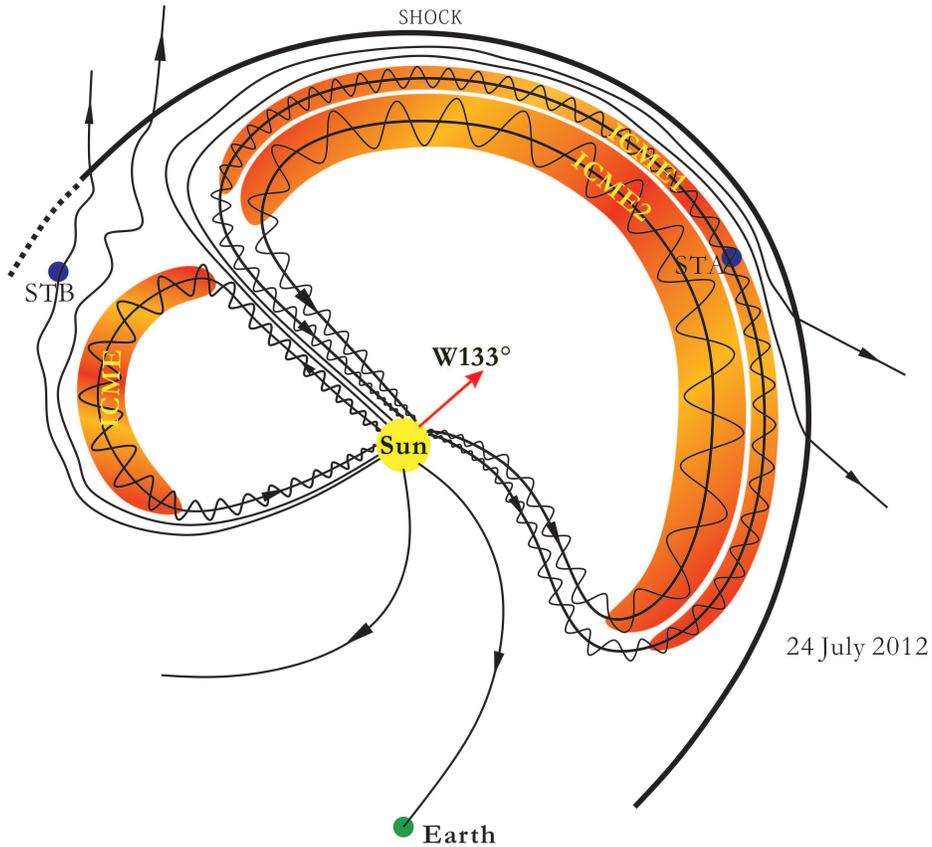}
\caption{Positions of both STEREO spacecraft (blue filled circles) and the Earth (green filled circle) in the ecliptic plane on 2012 July 24. 
Also shown are twin ICMEs which contributed to the extreme event observed at STEREO A and another ICME observed at STEREO B.
The black curve marks the shock observed at STEREO A and the black dashed curve marks the shock observed at STEREO B.
The thin arrowed black curves represent the interplanetary magnetic fields. 
The red arrow denotes the solar source longitude of the twin ICMEs.
The longitudinal extents and shapes of the shock and ICMEs are speculative for illustration purposes based on the in situ plasma and magnetic field measurements.}
\end{figure}

\begin{figure}[!htb]
\centering
\includegraphics[width=39pc]{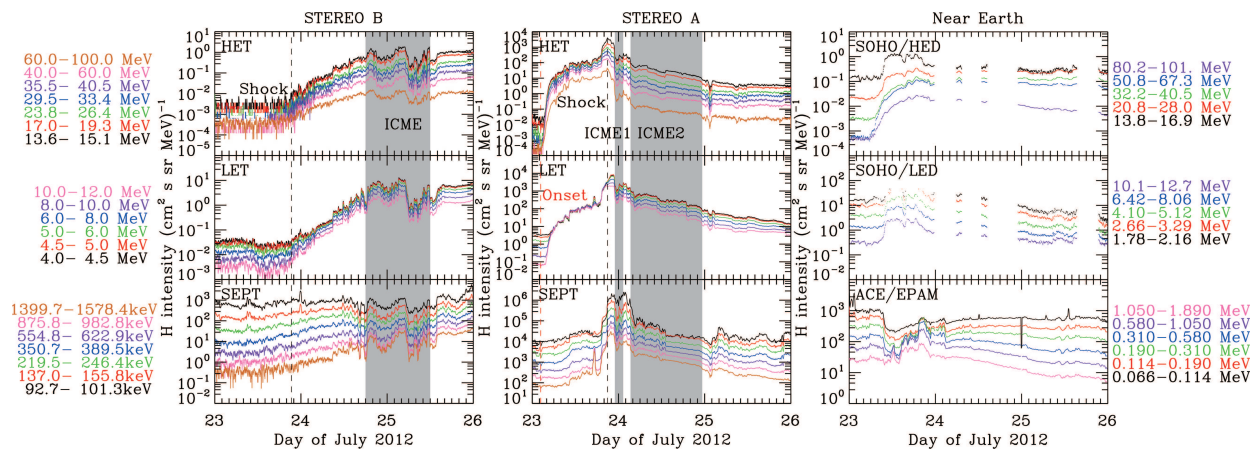}
\caption{Ten minute averaged proton intensities measured at STEREO B (left), A (middle) and near the Earth (right) in different energy channels. 
Dashed lines mark the time of the solar eruption (red) and the arrival times of the shocks (black).
Shadow regions indicate the ICME intervals. 
Date gaps are present in the SOHO data.}
\end{figure}
\clearpage

\begin{figure}[!htb]
\centering
\noindent\includegraphics[width=35pc]{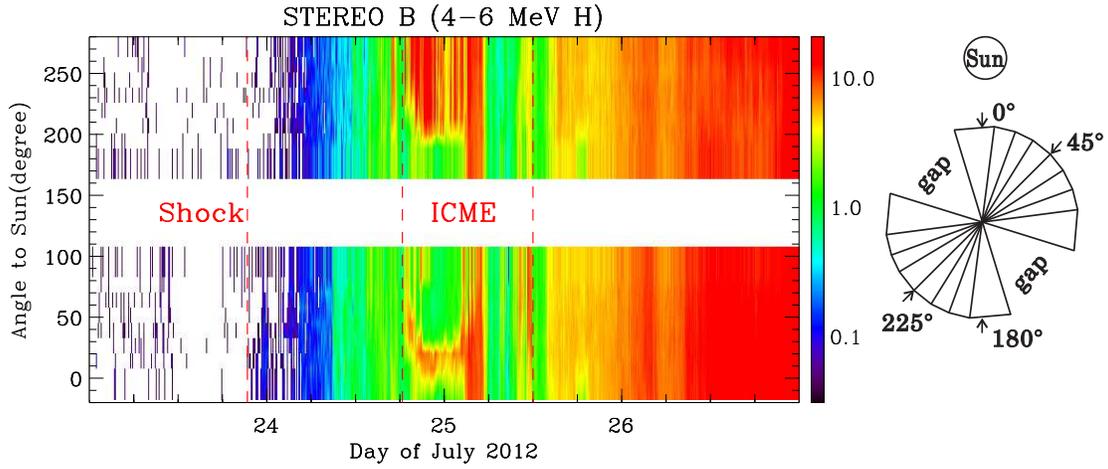}
\caption{Left panel: Intensity spectrogram of 4-6 MeV protons in each of the 16 sectors from LET on STEREO B during 2012 July 23-26,
The white band indicates one of the gaps in the LET field of view. 
The left red vertical dashed line marks the shock, and other two red lines indicate the boundaries of the ICME at STEREO B.
Right panel: Sketch of LET instrument with viewing directions in the ecliptic.
Each of the 16 sectors is represented by a different wedge. 
Particles coming in a straight line from the Sun would arrive at 0\degree${ }$ and those coming along the average Parker spiral angle would arrive at ${\sim}$45\degree .}
\end{figure}

\begin{figure}[!htb]
\centering
\noindent\includegraphics[width=28pc]{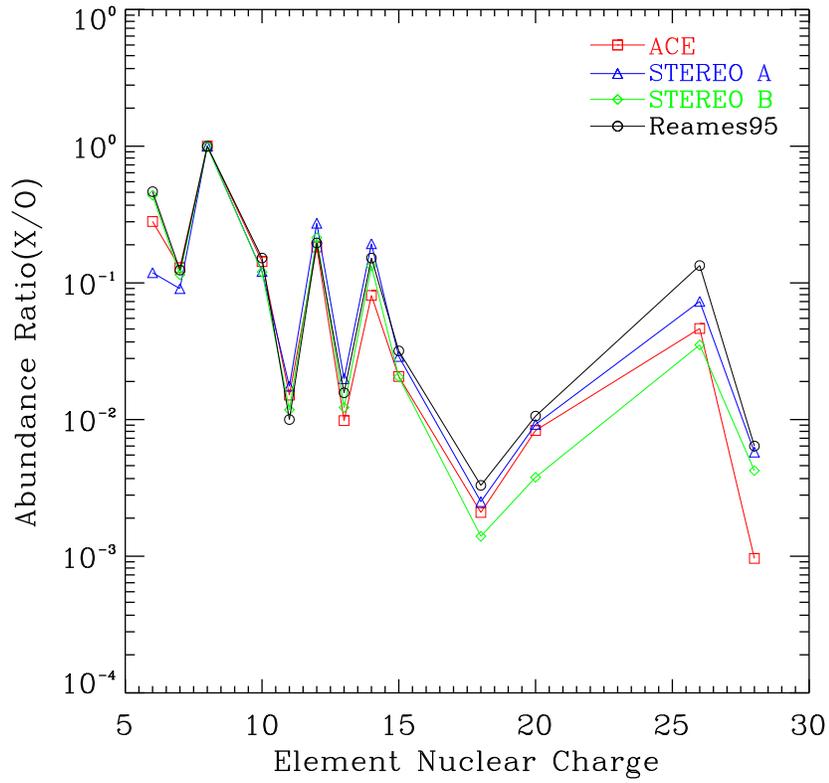}
\caption{Abundance ratios at 12-33 MeV nucleon$^{-1}$ (normalized to oxygen) as a function of atomic number, integrated over the event at each spacecraft.
The black line represents the average SEP abundances at 5-12 MeV as reported by \citet{Reames95}.}
\end{figure}

\begin{figure}[!htb]
\centering
\noindent\includegraphics[width=35pc]{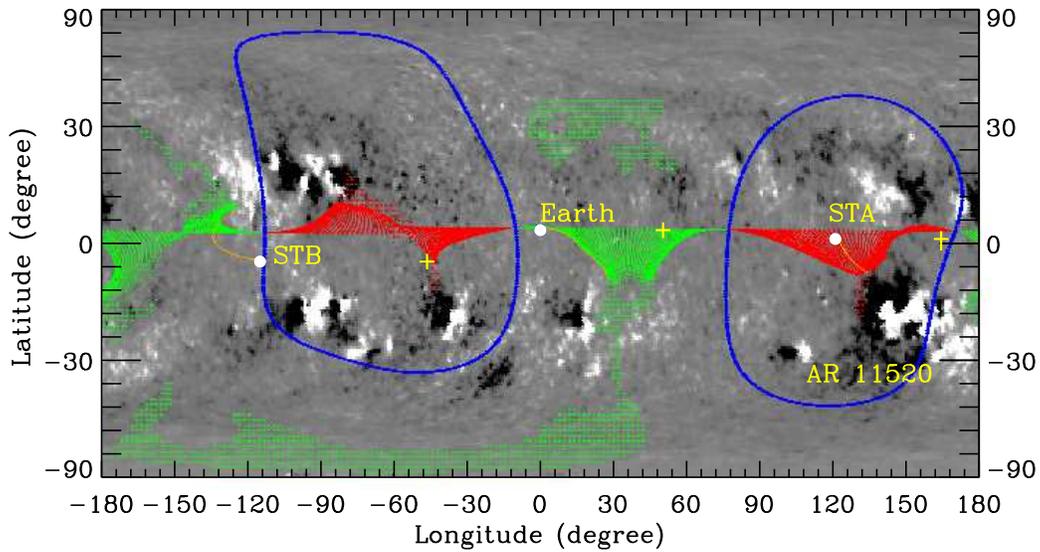}
\caption{GONG synoptic map with the PFSS open magnetic field lines connecting to the ecliptic plane. 
Red dots denote the negative polarity, and green dots mark the positive polarity.
The blue lines rounded as two wavy circles represent the current sheet separating different polarities.
The projections of both STEREO spacecraft and the Earth are indicated by the filled white circles, and the yellow crosses mark the corresponding longitudes of the three observers at 2.5 R$_{\odot}$.}
\end{figure}

\begin{figure}[!htb]
\centering
\noindent\includegraphics[width=25pc]{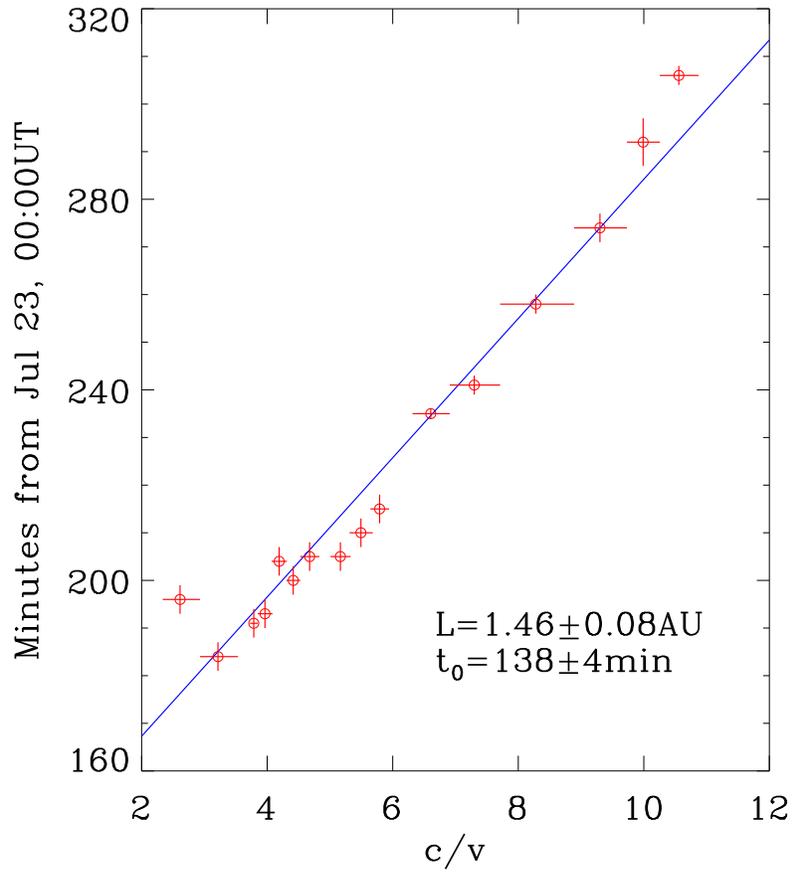}
\caption{Velocity dispersion analysis for STEREO A. The circles indicate the    {proton} onset times observed by HET and LET. The blue line is the linear fit to all points.
The field line length and particle release time on the Sun from the fit are also given.}
\end{figure}

\begin{figure}[!htb]
\centering 
\noindent\includegraphics[width=38pc]{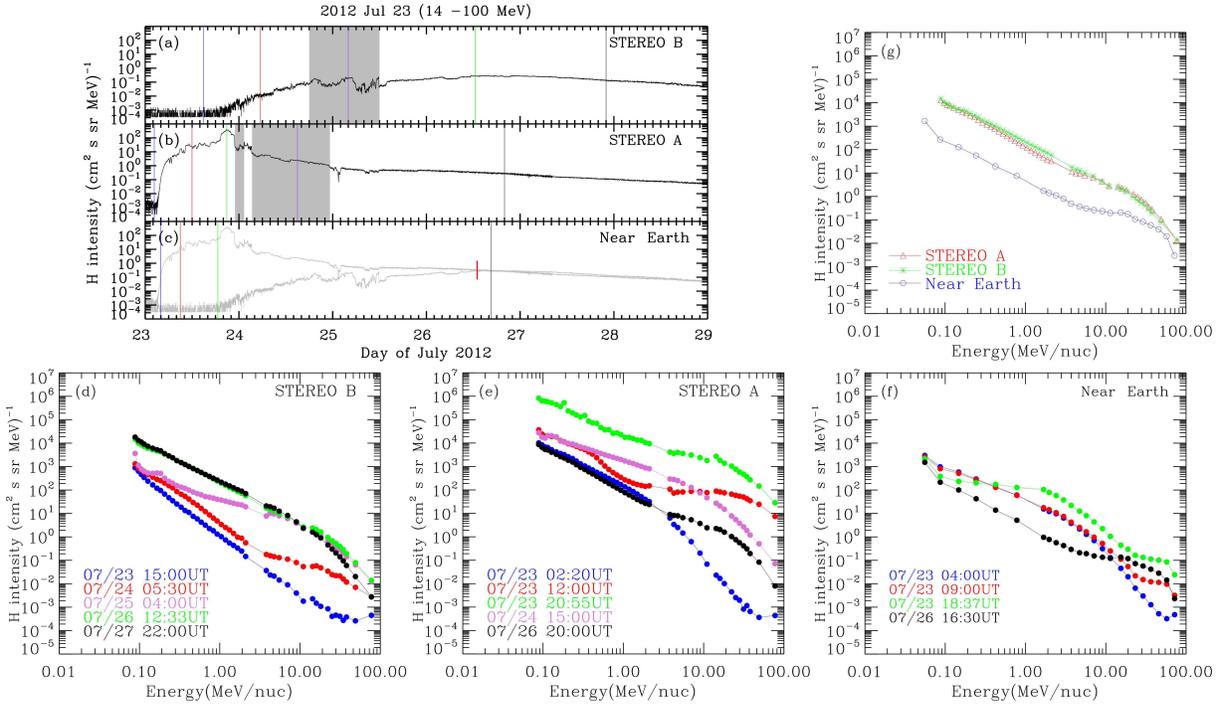}
\caption{a-c: Energetic proton (14-100 MeV) intensity profiles observed by HET on both STEREO and by ERNE on SOHO (taking into account the geometry factors of different energy bands). 
Panel (c) gives the proton intensity profile observed by SOHO along with a replotting of the STEREO proton intensities.
d-f: Average proton spectra over two-hour intervals around the vertical lines in panels (a-c). 
The same color is used for the line and corresponding spectrum.
g: Average proton spectra over two-hour intervals around the vertical red bar in panel (c).
The shaded regions in panels (a-c) indicate the ICME intervals.}
\end{figure}

\end{document}